\documentclass[prd,aps,floats,floatfix,amsmath,nofootinbib,superscriptaddress]{revtex4-1}
\usepackage{setspace}
\usepackage{axodraw4j}
\usepackage{pstricks}
\usepackage{graphicx}
\usepackage{bm}
\usepackage{epsfig}
\usepackage{latexsym}
\usepackage{color}
\usepackage{colordvi}
\newcommand{\nn}{\nonumber}

\newcommand{\beq}{\begin{equation}}
\newcommand{\eeq}{\end{equation}}
\newcommand{\bea}{\begin{eqnarray}}
\newcommand{\eea}{\end{eqnarray}}
\begin{document}


\title{Structure Function Sum rules for Systems with Large Scattering Lengths}

\author{Walter D.~Goldberger}
\affiliation{Department of Physics, Yale University,
    New Haven, CT 06520}
\author{Ira Z.~Rothstein}
\affiliation{Department of Physics, Carnegie Mellon University,
    Pittsburgh, PA 15213}

\begin{abstract}
We use a dispersion relation in conjunction with the operator product expansion (OPE) to derive model independent sum rules for  the dynamic structure functions of systems with large scattering lengths.   
We present an explicit sum rule for the structure functions  that control the density and spin response of the many-body ground state.   Our methods are general, and apply to either fermions or bosons which interact through two-body contact interactions with  large scattering lengths.  By employing a Borel transform of the OPE,  the relevant integrals are weighted towards infrared frequencies, thus allowing for greater overlap with low energy data.   Similar sum rules can be derived for other response functions. The sum rules can be used 
to extract the contact parameter introduced by Tan, including universality violating corrections at finite scattering lengths.
\end{abstract}

\maketitle


\section{Introduction}

The theory of non-relativistic spin-1/2 fermions near the unitarity limit (with $S$-wave scattering length $a\rightarrow\infty$), described by the Lagrangian
\beq
\label{eq:lag}
{\cal L} = \sum_{\alpha=\uparrow,\downarrow} \psi_\alpha^\dagger\left( i\partial_t + {\nabla^2\over 2m}\right)\psi_\alpha  + g(\psi_\uparrow^\dagger \psi_\downarrow^\dagger) (\psi_\uparrow  \psi_\downarrow),
\eeq
  plays a role in several areas of physics, ranging from nuclear physics to atomic and condensed matter physics.  Due to the recent experimental progress in manipulating gases of trapped cold atoms tuned to a Feshbach resonance, it has become a problem of relevance to obtain first principles predictions for the many-body collective behavior of fermionic systems near unitarity.    

Although the dynamics in the limit $a\rightarrow\infty$ is universal,  the lack of the textbook small expansion parameter, $na^3$, with $n$ the atomic number density, makes it difficult to derive model independent predictions for the interacting quantum many-body ground state of the system.   However, starting with recent work of Tan~\cite{tan1,tan2,tan3}, it has become apparent that many observables of the system are governed, in the $a\rightarrow\infty$ limit, by a set of universal relations which hold true in any quantum state of Eq.~(\ref{eq:lag}), and which can be fixed by studying the short distance/time scale behavior of the system.   These observables are summarized in the recent review~\cite{Braaten:2010if}, and besides those found in~\cite{tan1,tan2,tan3} include rf spectroscopy~\cite{rf,Braaten:2010dv}, photoassociation~\cite{photo,legget}, static~\cite{stat} and dynamic~\cite{st,tr} structure factors, and hydrodynamics~\cite{tr}.

From the point of view of quantum field theory, the existence of universal relations which hold in any quantum state are a consequence of of an enhanced non-relativistic conformal (Schrodinger~\cite{Hagen:1972pd,niederer,Mehen:1999nd,Nishida:2007pj}) symmetry in the ultraviolet (corresponding to the limit $a\rightarrow\infty$), together with the operator product expansion (OPE).  The OPE was originally introduced by Wilson~\cite{wilson} in the context of strong interaction physics, and the connection between the OPE in non-relativistic quantum field theory and universal relations such as those of~\cite{tan1,tan2,tan3} was first made explicit in the work of Braaten and Platter~\cite{bp,bkp}, which reproduced and extended the original Tan relations using the OPE.   More recently, refs.~\cite{Braaten:2010dv,st} showed that the OPE predicts that the asymptotic large frequency behavior of dynamic structure functions is also controlled by the Tan contact parameter ${\cal C}$.

By itself, the OPE only controls the properties of short distance/time observables, e.g., tails of distributions in the large frequency or momentum limit.  However, when used in conjunction with dispersion relations it is possible to obtain indirect information about low energy physics, in particular the spectrum of low energy excitations about the ground state, from the OPE.     In this paper, we will use the OPE in conjunction with a set of weighted (``Borel transformed") dispersion relations to derive sum rules that probe the infrared properties of  many particle systems whose dynamics is described by Eq.~(\ref{eq:lag}).     The sum rules relate weighted frequency integrals of retarded correlators to thermodynamic properties of the many-body ground state, in particular the number density, energy density, and the Tan contact ${\cal C}$.    These weighted integrals depend on a free parameter $\omega_0$ which, if taken small enough, are dominated by low energy data, and are therefore sensitive to the spectrum of excitations about the ground state.

The methods we employ to obtain these sum rules are based on analogous techniques used first used in a relativistic setting to compute QCD sum rules.~\cite{svz}.   Our sum rules are  applicable both to fermionic systems and to bosons near unitarity (e.g., He-4) and include correction to the infinite scattering length limit.
The techniques can be applied to any response function, but for the sake of illustration we will here
 consider only the density-density structure factor $S(\omega,{\vec q})$ .

In sec.~\ref{sec:opesr}, we review how the OPE controls the asymptotic behavior of Green's functions, and use this to obtain sum rules.   We apply this methodology to the structure function $S(\omega,{\vec q})$ in sec.~\ref{sec:nn}.   The main results are given in Eq.~(\ref{eq:res1}) and Eq.~(\ref{eq:res2}), and are valid for arbitrary values of the scattering length, but neglect subdominant effects in the OPE  from the range $r_0$ of the two-body potential or from local operators that mediate three-body physics and higher.   Taking the  $a\rightarrow\infty$ limit of our calculation, the results of this section agree with those of~\cite{st}.   In sec.~\ref{sec:pol}, we consider dynamic spin-density correlators, and find, as expected,  that the leading $\omega\rightarrow\infty$ asymptotics is entirely fixed in terms of the Tan contact ${\cal C}$.    The sum rules for the spin structure function derived in sec.~\ref{sec:pol} are of phenomenological interest since they allow for direct access to the contact ${\cal C}$ from low energy data.   Results for spinless bosons are qualitatively similar to those obtained for fermion in sec.~\ref{sec:nn}, and are given in sec.~\ref{sec:bose}.   Conclusions and directions for future work are discussed in sec.~\ref{sec:conc}

\section{Sum rules from the OPE}
\label{sec:opesr}
\subsection{The OPE and its range of validity}
In quantum field theory, the OPE is the statement that an operator product $A(x) B(0)$, in the limit of short distance/time scales, is equivalent to an expansion over local operators at $x=0$,
 \beq
A(x) B(0)\sim \sum_\alpha C_\alpha(x) {\cal O}_\alpha(0).
\eeq
This relation implied by this equation is strong in the sense that it holds valid inside all matrix elements, i.e. it is a statement about the operator algebra itself.  Inserted inside a matrix element or correlation function, the OPE relation is an asymptotic expansion \footnote{Exactly at the critical point, where the scattering length diverges, the expansion is in fact convergent.} in the limit where the distance scale $x$ is much shorter than the scales associated with external states or with additional operator insertions.

In a non-relativistic field theory (with $d=3$ spatial dimensions), the OPE takes the form
\beq
\label{eq:OPE}
A(x) B(0)\sim \sum_\alpha |{\vec x}|^{\Delta_\alpha-\Delta_A-\Delta_B} f_\alpha\left({|{\vec x}|^2\over t}\right) {\cal O}_\alpha(0),
\eeq
In this equation, $\Delta_{A,B,\alpha}$ are the scaling dimensions of the operators $A(x),B(x),{\cal O}(x)$ respectively, defined by 
\beq
{\cal O}(\lambda{\vec x},\lambda^2 t) = \lambda^{-\Delta} {\cal O}({\vec x},t),
\eeq
and the coefficient function $f_\alpha(|{\vec x}|^2/t)$ is a calculable function, as we will discuss in detail in sec.~\ref{sec:ope}.  Taking ground state matrix elements on both sides of Eq.~(\ref{eq:OPE}) (or ensemble averages at finite temperature $T$), one finds that the asymptotic short distance or time properties of two-point Green's functions, for instance the response functions of the theory, are dominated by the condensates $\langle {\cal O}_\alpha\rangle$ of the first few operators of lowest dimension $\Delta_\alpha$.

In this paper, we will use  Eq.~(\ref{eq:OPE}) to determine the asymptotic behavior of response functions and to derive sum rules for frequency dependent transport coefficients.    Ideally, one would like use the OPE to constrain the retarded Green's functions,
\beq
\label{eq:ret}
iG_R(q) = \int d^4 x e^{i q\cdot x} \theta(t)\langle [{\cal O}(x),{\cal O}(0)]\rangle,
\eeq
where $q^\mu=(\omega,{\vec q})$, and $\langle\cdots\rangle$ is either a vacuum expectation value (at $T=0$) or a thermodynamic average $\langle\cdots\rangle=\mbox{Tr}[e^{-\beta(H-\mu N)}\cdots]/{\cal Z}$.   In this equation the operator  ${\cal O}(x)$ is typically taken to be an operator corresponding to a conserved current (e.g., particle number or the energy-momentum tensor).   While Eq.~(\ref{eq:ret}) is the Green's function that is most closely related to experimental observables, in particular transport properties, it is more convenient in what follows to work with the Feynman (time ordered) function,
\beq
iG_F(q) = \int d^4 x e^{i q\cdot x} \langle T[{\cal O}(x) {\cal O}(0)]\rangle,
\eeq
which in the OPE limit takes the form
\beq
\label{eq:Fope}
iG_F(q)\sim \sum_\alpha {1\over\omega^{5/2+\Delta_\alpha/2-\Delta_{\cal O}}} c_\alpha\left({{\vec q}^2\over 2 m\omega}\right) \langle{\cal O}_\alpha(0)\rangle.
\eeq
The functions $c_\alpha(z)$, with $z={\vec q}^2/2m\omega$ are analytic functions of the variable $z$, except for single-particle poles at $z=1$ and branch cuts at values of $z$ corresponding to multi-particle states.  As pointed out in \cite{Braaten:2010if}, for the theory defined by Eq.~(\ref{eq:lag}), $c_\alpha(z)$ are \emph{exactly calculable functions} for all $z$ in the case where ${\cal O}_\alpha$ is in the one or two body sector.   If one is interested in the asymptotic behavior for $\omega\rightarrow\infty$ and ${\vec q}\rightarrow 0$ such that $z\rightarrow 0$, the functions $c_\alpha(z)$ can be expanded in a power series in $z$.

The condensates $\langle {\cal O}_\alpha\rangle$ are in general dependent on temperature $T$, chemical potential $\mu$ and the scattering length\footnote{In addition, $\langle {\cal O}_\alpha\rangle$ may depend upon other parameters characterizing the ground state, for instance the frequency $\omega_0$ of a harmonic trapping potential.} .    By dimensional analysis, the functional dependence is of the form
\beq	
\langle{\cal O}_\alpha\rangle = \mu^{\Delta_\alpha/2} g_\alpha\left({T\over\mu}, (a\sqrt{\mu})^{-1}\right),
\eeq
so that in the low temperature, $a\rightarrow\infty$ limit we can take $\langle{\cal O}_\alpha\rangle \sim \mu^{\Delta_\alpha/2} g_\alpha(0)$, and thus the $\omega\rightarrow\infty$ limit of a two-point correlator in the many-body state becomes an expansion in (non-analytic) powers of $\mu/\omega\ll 1$.   Strictly speaking, the OPE derived using Eq.~(\ref{eq:lag}) incorporates only the leading order (scattering length) two-body interaction induced by the effective range expansion of the $2\rightarrow 2$ scattering amplitude.    Thus observables calculated using Eq.~(\ref{eq:lag}) are accurate up to corrections parametrized by powers of $r_0\sqrt{2 m \omega},$ where $r_0$ is a length scale characterizing the range of the two-body potential (e.g., the van der Waals interaction length scale $\ell_{vDW}$).   This means that our results for the OPE are valid in the range of energies
\beq
\mu\ll \omega \ll {1\over 2 m r_0^2}.
\eeq   
In this window, the OPE implies the asymptotic ($\omega\rightarrow\infty,{\vec q}\rightarrow 0$) expansion
\beq
iG_F(q)\sim {1\over \omega^{5/2-\Delta_{\cal O}}}\sum_{\alpha,i} b_{\alpha,i} \left({\mu\over\omega}\right)^{\Delta_\alpha/2} \left({{\vec q}^2\over 2m\omega}\right)^{p_{\alpha,i}},
\eeq
where $b_{\alpha,i}$ are numerical coefficients proportional to $\langle{\cal O}_\alpha\rangle$, and $p_{\alpha,i}$ are integer.

\subsection{Sum Rules}

The asymptotic behavior of the Feynman Green's function implied by the OPE, together with analyticity/causality properties of the retarded two-point function, can be used to obtain model independent sum rules that must be satisfied by the dynamical response functions of the system.   
It is these analytic properties that will allow us to get useful information about the
small frequency behavior of the system from the OPE.

The starting point for this is the Kramers-Kronig dispersion relation satisfied by $G_R(q)$.    Because $G_R(\omega,{\vec q})$ is an analytic function in the upper-half complex $\omega$ plane $\mbox{Im}(\omega)>0$, it follows that 
\bea
\label{eq:disp}
\nonumber
\mbox{Re } G_R(\omega) &=& \mbox{Pr } \int_0^\infty {d\omega'^2\over \pi} {\mbox{Im } G_R(\omega')\over {\omega'}^2 - \omega^2},\\
\mbox{Im } G_R(\omega) &=& -2 \omega \mbox{Pr } \int_0^\infty {d\omega'\over \pi} {\mbox{Re } G_R(\omega')\over {\omega'}^2 - \omega^2}.
\eea
These expressions are strictly valid if the function $G_R(\omega)$ decays sufficiently rapidly for $\omega\rightarrow\infty$.   Otherwise, further subtraction terms are needed.   These introduce additional free parameters and therefore diminish the predictive power of the dispersion relations.

Assuming naively that  $G_R(\omega)$ decays rapidly enough, it is possible to expand both sides of the dispersion relation to obtain,
\beq
\label{eq:naive}
\mbox{Re } G_R(\omega\rightarrow\infty) =  -\sum_{n=0}^\infty {1\over (\omega^2)^{n+1}} \int_0^\infty {d\omega'^2\over \pi} {\omega'}^{2n} {\mbox{Im } G_R(\omega')}.
\eeq
It follows from the spectral representation of the two-point functions that, for real $\omega$,  $\mbox{Re } G_R(\omega)=\mbox{Re } G_F(\omega)$.   It is therefore legitimate to replace the LHS of this equation with Eq.~(\ref{eq:Fope}).      If the asymptotic expansion of $\mbox{Re } G_F(\omega\rightarrow\infty)$ on the real axis consisted of analytic powers of $1/\omega^2,$ it would then be possible to use Eqs.~(\ref{eq:Fope},\ref{eq:naive}) to obtain simple expressions for the moments
\beq
I_n =\int_0^\infty {d\omega^2\over \pi} {\omega}^{2n} {\mbox{Im } G_R(\omega')},
\eeq

\noindent in terms of the condensates $\langle {\cal O}_\alpha\rangle$.   Indeed, this is a standard textbook procedure 
for deriving sum rules.    While this procedure does lead to several useful moment sum rules (the $f$-sum rule being among them), it fails but for the first few moments $I_n$.   The problem is that the asymptotic decay of correlators in any quantum field theory, as predicted by the OPE, is at best power law, and the dispersion integrals in Eq.~(\ref{eq:disp}) are not convergent.    

In order to sidestep this problem, we follow instead the procedure originally used in~\cite{svz} to derive similar sum rules in QCD.   Consider the contour integral
\beq
I=\oint {dz\over 2\pi i} {z \, G_R(z)\over z^2 + \omega^2},
\eeq
taken over a contour consisting of the real axis plus a semi-circular arc of radius $R\rightarrow\infty$ through the upper-half plane containing the point $z=i\omega$ ($\mbox{Im} \omega =0,\omega>0$).   Since $\mbox{Re }G_R(-z) = \mbox{Re }G_R(z)$, and $\mbox{Im }G_R(-z) =- \mbox{Im }G_R(z)$, for real $z$, the residue theorem gives
\beq
\int_0^\infty {d{\omega'}^2\over \pi} { \mbox{Im } G_R(\omega')\over {\omega'}^2 + \omega^2} = G_R(i\omega) =G_F(i\omega),
\eeq
where we have used the fact that $G_R(\omega)=G_F(\omega)$ for complex $\omega$, which follows by comparing the spectral representations.    The spectral representation for $G_F(\omega)$ also implies that $G_F(i\omega)\equiv G_F(-\omega^2)$ is a real function for real $\omega$, and thus by successive differentiation, we obtain 
\beq
\int_0^\infty {d{\omega'}^2\over \pi} { \mbox{Im } G_R(\omega')\over ({\omega'}^2 + \omega^2)^{n+1}} = {1\over n!} \left(-{d\over d\omega^2}\right)^{n} G_F(-\omega^2).
\eeq
Combining this with the OPE of Eq.~(\ref{eq:Fope}), written in the schematic form 
\beq
G_F(-\omega^2) = \sum_\alpha a_\alpha({\vec q}) \left(\omega^2\right)^{-\alpha},
\eeq
with $a_\alpha({\vec q})$ fixed in terms of the condensates $\langle {\cal O}_\alpha\rangle$, the sum rules become
\beq
\label{eq:sr}
\int_0^\infty {d{\omega'}^2\over \pi} { \mbox{Im } G_R(\omega')\over ({\omega'}^2 + \omega^2)^{n+1}} = \left({1\over\omega^2}\right)^n \sum_\alpha {\Gamma(\alpha+n)\over \Gamma(\alpha)\Gamma(n+1)} a_\alpha({\vec q}) (\omega^2)^{-\alpha}.
\eeq
This is a sum rule that relates thermodynamic averages on the RHS (the condensates $\langle{\cal O}_\alpha\rangle$) to the function $\mbox{Im }G_R(\omega)$ on the LHS.    By standard Kubo formulae, $\mbox{Im }G_R(\omega)$ controls the dissipative response to perturbations generated by the source ${\cal O}$ and can be measured experimentally (in principle).  Thus the result in Eq.~(\ref{eq:sr}) can be either tested against experimental results, or used to constrain the parameters of models that make specific predictions for $\mbox{Im }G_R(\omega)$.

For sufficiently low $n$, the integrals on the LHS of this formula may still fail to converge.    However, it is possible to take the limit $n\rightarrow\infty$, with $\omega_0^2\equiv \omega^2/n$ fixed on both sides of this equation, which yields
\beq
\label{eq:borelsr}
{1\over \omega_0^2}\int_0^\infty {d\omega^2 \over\pi} e^{-\omega^2/\omega_0^2} \mbox{Im} G_R(\omega) = \sum_\alpha {a_{\alpha}({\vec q})\over\Gamma(\alpha)} (\omega_0^2)^{-\alpha}.
\eeq
This result should be of phenomenological relevance, provided it is possible to choose $\omega_0$ large enough so that the OPE is valid, yet small enough so that the integral on the LHS is still dominated by the low energy data.

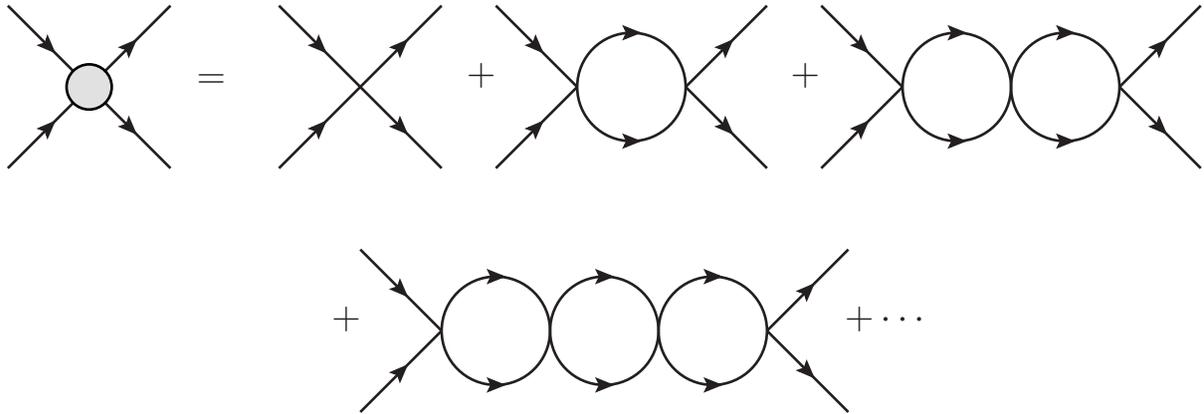
\begin{figure*}[t!]
\begin{center}
\fcolorbox{white}{white}{
  \begin{picture}(451,154) (26,-26)
    \SetWidth{0.5}
    \SetColor{Black}
    \Text(98.77,97.067)[lb]{\Large{\Black{$=$}}}
    \Text(200.945,97.067)[lb]{\Large{\Black{$+$}}}
    \Text(323.556,97.067)[lb]{\Large{\Black{$+$}}}
    \Text(343.991,5.109)[lb]{\Large{\Black{$+\cdots$}}}
    \SetWidth{1.0}
    \Line[arrow,arrowpos=0.5,arrowlength=5,arrowwidth=2,arrowinset=0.2](27.247,127.72)(57.9,97.067)
    \Line[arrow,arrowpos=0.5,arrowlength=5,arrowwidth=2,arrowinset=0.2](27.247,66.414)(57.9,97.067)
    \Line[arrow,arrowpos=0.5,arrowlength=5,arrowwidth=2,arrowinset=0.2](57.9,97.067)(88.552,66.414)
    \Line[arrow,arrowpos=0.5,arrowlength=5,arrowwidth=2,arrowinset=0.2](57.9,97.067)(88.552,127.72)
    \GOval(57.9,97.067)(8.515,8.515)(0){0.882}
    \Line[arrow,arrowpos=0.5,arrowlength=5,arrowwidth=2,arrowinset=0.2](129.422,127.72)(160.075,97.067)
    \Line[arrow,arrowpos=0.5,arrowlength=5,arrowwidth=2,arrowinset=0.2](129.422,66.414)(160.075,97.067)
    \Line[arrow,arrowpos=0.5,arrowlength=5,arrowwidth=2,arrowinset=0.2](160.075,97.067)(190.728,127.72)
    \Line[arrow,arrowpos=0.5,arrowlength=5,arrowwidth=2,arrowinset=0.2](160.075,97.067)(190.728,66.414)
    \Line[arrow,arrowpos=0.5,arrowlength=5,arrowwidth=2,arrowinset=0.2](211.163,127.72)(241.816,97.067)
    \Line[arrow,arrowpos=0.5,arrowlength=5,arrowwidth=2,arrowinset=0.2](211.163,66.414)(241.816,97.067)
    \Arc[arrow,arrowpos=0.5,arrowlength=5,arrowwidth=2,arrowinset=0.2,clock](262.251,97.067)(20.435,-180,-360)
    \Arc[arrow,arrowpos=0.5,arrowlength=5,arrowwidth=2,arrowinset=0.2](262.251,97.067)(20.435,-180,0)
    \Line[arrow,arrowpos=0.5,arrowlength=5,arrowwidth=2,arrowinset=0.2](282.686,97.067)(313.339,66.414)
    \Line[arrow,arrowpos=0.5,arrowlength=5,arrowwidth=2,arrowinset=0.2](282.686,97.067)(313.339,127.72)
    \Line[arrow,arrowpos=0.5,arrowlength=5,arrowwidth=2,arrowinset=0.2](333.774,66.414)(364.426,97.067)
    \Line[arrow,arrowpos=0.5,arrowlength=5,arrowwidth=2,arrowinset=0.2](333.774,127.72)(364.426,97.067)
    \Arc[arrow,arrowpos=0.5,arrowlength=5,arrowwidth=2,arrowinset=0.2,clock](384.861,97.067)(20.435,-180,-360)
    \Arc[arrow,arrowpos=0.5,arrowlength=5,arrowwidth=2,arrowinset=0.2](384.861,97.067)(20.435,-180,0)
    \Arc[arrow,arrowpos=0.5,arrowlength=5,arrowwidth=2,arrowinset=0.2,clock](425.732,97.067)(20.435,-180,-360)
    \Arc[arrow,arrowpos=0.5,arrowlength=5,arrowwidth=2,arrowinset=0.2](425.732,97.067)(20.435,-180,0)
    \Line[arrow,arrowpos=0.5,arrowlength=5,arrowwidth=2,arrowinset=0.2](446.167,97.067)(476.819,66.414)
    \Line[arrow,arrowpos=0.5,arrowlength=5,arrowwidth=2,arrowinset=0.2](446.167,97.067)(476.819,127.72)
    \Line[arrow,arrowpos=0.5,arrowlength=5,arrowwidth=2,arrowinset=0.2](160.075,35.761)(190.728,5.109)
    \Line[arrow,arrowpos=0.5,arrowlength=5,arrowwidth=2,arrowinset=0.2](160.075,-25.544)(190.728,5.109)
    \Arc[arrow,arrowpos=0.5,arrowlength=5,arrowwidth=2,arrowinset=0.2,clock](211.163,5.109)(20.435,-180,-360)
    \Arc[arrow,arrowpos=0.5,arrowlength=5,arrowwidth=2,arrowinset=0.2](211.163,5.109)(20.435,-180,0)
    \Arc[arrow,arrowpos=0.5,arrowlength=5,arrowwidth=2,arrowinset=0.2,clock](252.033,5.109)(20.435,-180,-360)
    \Arc[arrow,arrowpos=0.5,arrowlength=5,arrowwidth=2,arrowinset=0.2](252.033,5.109)(20.435,-180,0)
    \Arc[arrow,arrowpos=0.5,arrowlength=5,arrowwidth=2,arrowinset=0.2,clock](292.903,5.109)(20.435,-180,-360)
    \Arc[arrow,arrowpos=0.5,arrowlength=5,arrowwidth=2,arrowinset=0.2](292.903,5.109)(20.435,-180,0)
    \Line[arrow,arrowpos=0.5,arrowlength=5,arrowwidth=2,arrowinset=0.2](313.339,5.109)(343.991,35.761)
    \Line[arrow,arrowpos=0.5,arrowlength=5,arrowwidth=2,arrowinset=0.2](313.339,5.109)(343.991,-25.544)
    \Text(149.858,5.109)[lb]{\Large{\Black{$+$}}}
  \end{picture}
}
\end{center}
\caption{The exact $2\rightarrow 2$ scattering amplitude.\label{fig:eamp}}
\end{figure*}

\section{Asymptotics of the Dynamic Structure Factor}
\label{sec:nn}

\label{sec:ope}

We know apply the general results of the previous section to the case of the asymptotic structure factor $S(\omega,{\vec q})$, which is related to the retarded correlator of the density\footnote{In what follows, we employ the convention that suppressed spin indices are summed over.   E.g., $\psi^\dagger \psi = \sum_{\alpha=\uparrow,\downarrow} \psi^\dagger_\alpha \psi_\alpha,$ etc.} operator ${\cal O}(x) =n(x)\equiv [\psi^\dagger\psi](x)$,
\beq
S(\omega,{\vec q}) = -{1\over \pi}\mbox{Im } G_R(\omega,{\vec q}),
\eeq
at real frequency $\omega>0$.  The asymptotic behavior of this correlator for fermions strictly at the unitarity limit has been studied previously in~\cite{st} using the OPE, and by~\cite{tr} using the methods of ref.~\cite{legget}.

Here we will determine the structure of the OPE for 
\beq
\label{eq:nnope}
M(q)\equiv\int d^4 x e^{i q\cdot x} T \left[n(x) n(0)\right]\sim\sum_\alpha C_\alpha(q) {\cal O}_\alpha(0).
\eeq
Since the OPE holds inside all matrix elements, one may calculate the low-lying Wilson coefficients $C_\alpha(q)$ by evaluating both sides of the above expression between states of small particle number, without reference to the interacting many-body ground state of the system.  Once the coefficients are fixed in this way, the OPE may be evaluated in the many-body ground state at finite $T$ and $\mu$ (or for polarized samples with differing chemical potentials $\mu_\uparrow,\mu_\downarrow$ for the two spin states).

We will compute the Wilson coefficients in Eq.~(\ref{eq:nnope}) up to dimension $\Delta=5$ (in $d=3$ spatial dimensions).   In particular, this excludes operators built out of six powers or more powers of $\psi,\psi^\dagger$ (by the particle number symmetry $\psi(x)\rightarrow e^{i\alpha} \psi(x)$, operators with odd powers cannot appear).   This implies that to fix the relevant coefficients, it is sufficient to compute matrix elements of Eq.~(\ref{eq:nnope}) between one-particle and two-particle states.    In the $1\rightarrow 1'$ and $2\rightarrow 2'$ sectors, the theory in Eq.~(\ref{eq:lag}) is simple enough that the coefficients $C_\alpha(q)$ are calculable \emph{exactly}.   

In order to compute the relevant matrix elements, we will use dimensional regularization to regulate infrared (IR) and ultraviolet (UV) divergences that arise at intermediate stages in the calculations.   Our results for the Wilson coefficients below will be manifestly IR finite, as well as UV finite after introducing suitably renormalized operators.

The basic element in the calculation is the exact amplitude for $2\rightarrow 2$ scattering, which is given by the sum over the graphs in Fig.~\ref{fig:eamp}.    The bubble expansion in Fig.~\ref{fig:eamp} is a geometric series which sums to 
\beq
{\cal A}^{-1}(\omega,{\vec q}) = -{1\over g} - f(\omega,{\vec q}),
\eeq
where $(\omega,\vec q)$ is the total energy/momentum of the incoming particles, and
\beq
\label{eq:func}
 f(\omega,{\vec q}) = {m \over (4\pi)^{d/2}}\Gamma(1-d/2)\left[-m(\omega-E_{\vec q}/2+i\epsilon)\right]^{d/2-1},
\eeq  
is the result of computing a one-loop Feynman diagram.   In general, this one-loop graph is UV divergent, but in $d>2$ this divergence is a power divergence that dimensional regularization sets to zero.   For example, in $d=3$ where the one-loop amplitude is linearly divergent, the above result becomes, in the center-of-mass frame,
\beq
{\cal A} = -{4\pi\over m}\cdot {1\over 1/a + i\sqrt{m\omega}},
\eeq
where the scattering length is
\beq
a= {mg\over 4\pi},
\eeq
so that in dimensional regularization, the unitary limit corresponds to $g\rightarrow\infty$.

\subsection{Matching in the one-particle sector}

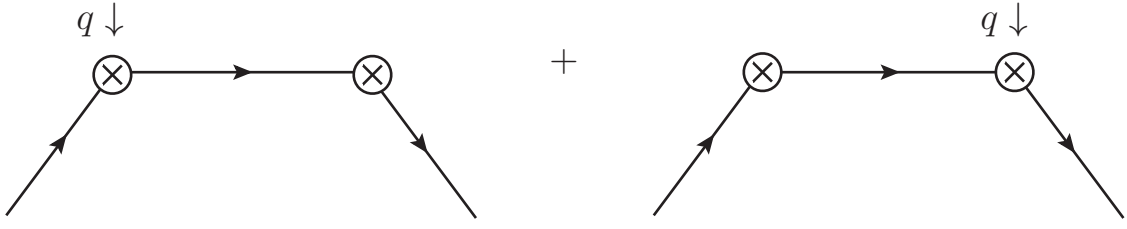
\begin{figure*}[t]
\begin{center}
\fcolorbox{white}{white}{
  \begin{picture}(423,90) (127,-168)
    \SetWidth{1.0}
    \SetColor{Black}
    \Line[arrow,arrowpos=0.5,arrowlength=5,arrowwidth=2,arrowinset=0.2](169,-112)(265,-112)
    \Line[arrow,arrowpos=0.5,arrowlength=5,arrowwidth=2,arrowinset=0.2](508,-112)(549,-167)
    \Line[arrow,arrowpos=0.5,arrowlength=5,arrowwidth=2,arrowinset=0.2](413,-112)(509,-112)
    \Line[arrow,arrowpos=0.5,arrowlength=5,arrowwidth=2,arrowinset=0.2](128,-166)(169,-111)
    \Line[arrow,arrowpos=0.5,arrowlength=5,arrowwidth=2,arrowinset=0.2](264,-112)(305,-167)
    \COval(266,-113)(7.071,7.071)(45.0){Black}{White}\Line(262.464,-116.536)(269.536,-109.464)\Line(262.464,-109.464)(269.536,-116.536)
    \COval(168,-113)(7.071,7.071)(45.0){Black}{White}\Line(164.464,-116.536)(171.536,-109.464)\Line(164.464,-109.464)(171.536,-116.536)
    \Text(334,-113)[lb]{\Large{\Black{$+$}}}
    \Text(155,-99)[lb]{\Large{\Black{$q\downarrow$}}}
    \Text(497,-99)[lb]{\Large{\Black{$q\downarrow$}}}
    \Line[arrow,arrowpos=0.5,arrowlength=5,arrowwidth=2,arrowinset=0.2](372,-166)(413,-111)
    \COval(413,-112)(7.071,7.071)(45.0){Black}{White}\Line(409.464,-115.536)(416.536,-108.464)\Line(409.464,-108.464)(416.536,-115.536)
    \COval(508,-112)(7.071,7.071)(45.0){Black}{White}\Line(504.464,-115.536)(511.536,-108.464)\Line(504.464,-108.464)(511.536,-115.536)
  \end{picture}
  }
\caption{Diagrams contributing to the matrix element  of the operator $M(q)$ between one-particle states.   The symbol $\otimes$ denotes an insertion of the operator $\psi^\dagger\psi$. \label{fig:1to1}}
\end{center}
\end{figure*}

The matrix element of the RHS of Eq.~(\ref{eq:nnope}) between one-particle states with momentum $p^\mu=(p^0,{\vec p})$  is given by the graphs in Fig.~\ref{fig:1to1}.   The result, which is exact to all orders in the coupling $g$, is given by
\beq
\langle p'|    M(q)   |p \rangle ={i\over \omega + p^0 - E_{\vec q+\vec p}} + {i\over -\omega + {p'}^0 - E_{\vec {p'}-\vec{q}}}.
\eeq
For reasons that will become clear below, we keep the one-particle states off-shell, meaning that we do not assume $p^0=E_{\vec p}\equiv {\vec p}^2/2m$.   We have suppressed the spin labels on the external states in this calculation, but note that the matrix element is zero unless initial and final spins are equal.

In order to match onto the OPE prediction, we take the limit $q\rightarrow\infty$ in this expression and expand the non-relativistic propagators.   Away from the one-particle thresholds $q^0=\omega =\pm E_{\vec q}$, the expansion is reproduced by the $1\rightarrow 1$ matrix elements of the operator $n(0)=\psi^\dagger\psi$ at dimension $\Delta=3$, the $\Delta=4$ operators
\bea
\nn
& \partial_i n , &\\
& {\vec J}_i = -{i\over 2m} \psi^\dagger \stackrel{\leftrightarrow}{\partial}_i\psi,&
\eea
and the $\Delta=5$ operators
\bea
\nn
 &\partial_i\partial_j n(0),  \partial_i J_j(0) + \partial_jJ_i(0),& \\ 
 \nn
  &{1\over 2m} \psi^\dagger  \stackrel{\leftrightarrow}{\partial}_i \stackrel{\leftrightarrow}{\partial}_j\psi,&\\
&\psi^\dagger \left(i\partial_t + {\nabla^2\over 2m}\right) \psi, \psi^\dagger {\Big(-i\stackrel{\leftarrow}{\partial}_t + {\stackrel{\leftarrow}{\nabla}^2\over 2m}\Big)} \psi.
\eea
The OPE in the one-body sector thus reads
\begin{widetext}
\bea
\label{eq:1part}
\nonumber
\left. M(q) \right|_{1\rightarrow 1'} &\sim& C_n(q) n(0) + C^{(1)}_i(q) \partial_i n(0)+ C^{(2)}_i(q) J_i(0) + C^{{(1)}}_{ij}(q) \partial_i\partial_j n(0)\\
\nn
& & {} + C^{{(2)}}_{ij}(q) \left[\partial_i J_j(0) + \partial_jJ_i(0)\right] + C^{(3)}_{ij}(q) {1\over 2m} \psi^\dagger  \stackrel{\leftrightarrow}{\partial}_i \stackrel{\leftrightarrow}{\partial}_j\psi\\
& & + C^{{(4)}}(q) \psi^\dagger \left(i\partial_t + {\nabla^2\over 2m}\right)\psi  + C^{(5)}(q) \psi^\dagger {\Big(-i\stackrel{\leftarrow}{\partial}_t + {\stackrel{\leftarrow}{\nabla}^2\over 2m}\Big)} \psi,
\eea
\end{widetext}

with,
\bea
\label{eq:cn}
C_n(q)               &=& {2iE_{\vec q}\over (\omega - E_{\vec q}) (\omega + E_{\vec q})},\\
C^{(1)}_i(q)       &=&  -{(\omega^2+ E^2_{\vec q})\over (\omega - E_{\vec q})^2 (\omega + E_{\vec q})^2} {{\vec q}_i\over m},\\
C^{(2)}_i(q)       &=&  -{4i\omega E_{\vec q} \over (\omega - E_{\vec q})^2 (\omega + E_{\vec q})^2} {\vec q}_i\\
C^{(1)}_{ij}(q)   &=&  -i {{\vec q}_i  {\vec q}_j\over 4 m^2} \left[{1\over (\omega - E_{\vec q})^3} - {1\over (\omega +E_{\vec q})^3}\right],\\
C^{(2)}_{ij}(q)   &=&  {{\vec q}_i  {\vec q}_j\over 2 m} \left[{1\over (\omega - E_{\vec q})^3} +{1\over (\omega +E_{\vec q})^3}\right],\\
C^{(3)}_{ij}(q)   &=&  -i {{\vec q}_i  {\vec q}_j\over 2 m} \left[{1\over (\omega - E_{\vec q})^3} -{1\over (\omega +E_{\vec q})^3}\right],\\
C^{(4)}(q) &=& -{i\over (\omega - E_{\vec q})^2},\\
\label{eq:c5}
C^{(5)}(q) &=& -{i\over (\omega + E_{\vec q})^2}.
\eea
Note that by the Heisenberg equations of motion for the field operator $\psi(x)$,
\beq
\psi^\dagger  \left(i\partial_t + {\nabla^2\over 2m}\right)\psi = -2g (\psi_\uparrow^\dagger \psi_\downarrow^\dagger) (\psi_\uparrow  \psi_\downarrow)
\eeq
However, because we are matching to the OPE coefficients using off-shell states, it is not permissible to apply equations of motion inside matrix elements at this point in the calculation.   The reason for matching to the OPE using off-shell states will become clear below, when we compute the terms in the OPE that are necessary to reproduce the $2\rightarrow 2'$ matrix elements of $M(q)$.


\subsection{Matching in the two-particle sector}

\begin{figure*}[t]
\begin{center}
\fcolorbox{white}{white}{
  \begin{picture}(386,108) (79,-31)
    \SetWidth{0.5}
    \SetColor{Black}
    \Text(119,-36)[lb]{\Large{\Black{$(a)$}}}
    \Text(263,-36)[lb]{\Large{\Black{$(b)$}}}
    \Text(407,-36)[lb]{\Large{\Black{$(c)$}}}
    \SetWidth{1.0}
    \Line[arrow,arrowpos=0.5,arrowlength=5,arrowwidth=2,arrowinset=0.2](80,76)(128,28)
    \Line[arrow,arrowpos=0.5,arrowlength=5,arrowwidth=2,arrowinset=0.2](80,-20)(128,28)
    \Line[arrow,arrowpos=0.5,arrowlength=5,arrowwidth=2,arrowinset=0.2](128,28)(176,76)
    \Line[arrow,arrowpos=0.5,arrowlength=5,arrowwidth=2,arrowinset=0.2](128,28)(176,-20)
    \GOval(128,28)(13,13)(0){0.882}
    \COval(160,-4)(4.243,4.243)(45.0){Black}{White}\Line(157.879,-6.121)(162.121,-1.879)\Line(157.879,-1.879)(162.121,-6.121)
    \COval(96,60)(4.243,4.243)(45.0){Black}{White}\Line(93.879,57.879)(98.121,62.121)\Line(93.879,62.121)(98.121,57.879)
    \Line[arrow,arrowpos=0.5,arrowlength=5,arrowwidth=2,arrowinset=0.2](224,76)(272,28)
    \Line[arrow,arrowpos=0.5,arrowlength=5,arrowwidth=2,arrowinset=0.2](224,-20)(272,28)
    \Line[arrow,arrowpos=0.5,arrowlength=5,arrowwidth=2,arrowinset=0.2](272,28)(320,-20)
    \Line[arrow,arrowpos=0.5,arrowlength=5,arrowwidth=2,arrowinset=0.2](272,28)(320,76)
    \GOval(272,28)(13,13)(0){0.882}
    \COval(240,60)(4.243,4.243)(45.0){Black}{White}\Line(237.879,57.879)(242.121,62.121)\Line(237.879,62.121)(242.121,57.879)
    \COval(240,-4)(4.243,4.243)(45.0){Black}{White}\Line(237.879,-6.121)(242.121,-1.879)\Line(237.879,-1.879)(242.121,-6.121)
    \Line[arrow,arrowpos=0.5,arrowlength=5,arrowwidth=2,arrowinset=0.2](368,76)(416,28)
    \Line[arrow,arrowpos=0.5,arrowlength=5,arrowwidth=2,arrowinset=0.2](368,-20)(416,28)
    \Line[arrow,arrowpos=0.5,arrowlength=5,arrowwidth=2,arrowinset=0.2](416,28)(464,-20)
    \Line[arrow,arrowpos=0.5,arrowlength=5,arrowwidth=2,arrowinset=0.2](416,28)(464,76)
    \GOval(416,28)(13,13)(0){0.882}
    \COval(384,60)(4.243,4.243)(45.0){Black}{White}\Line(381.879,57.879)(386.121,62.121)\Line(381.879,62.121)(386.121,57.879)
    \COval(400,44)(4.243,4.243)(45.0){Black}{White}\Line(397.879,41.879)(402.121,46.121)\Line(397.879,46.121)(402.121,41.879)
  \end{picture}
}
\caption{All tree graph topologies contributing to the the $2 \rightarrow 2$ matrix element of the operator $M(q)$.   The shaded blobs denote exact vertices, given in Fig.~\ref{fig:eamp}.  The symbol $\otimes$ denotes an insertion of the operator $\psi^\dagger\psi$.  \label{fig:tree}}
\end{center}
\end{figure*}
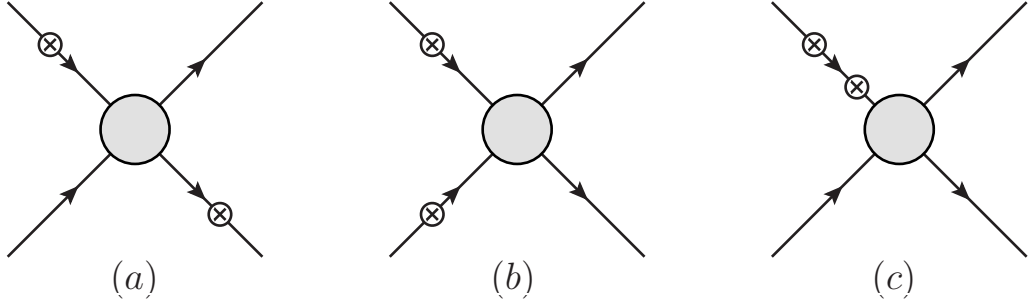

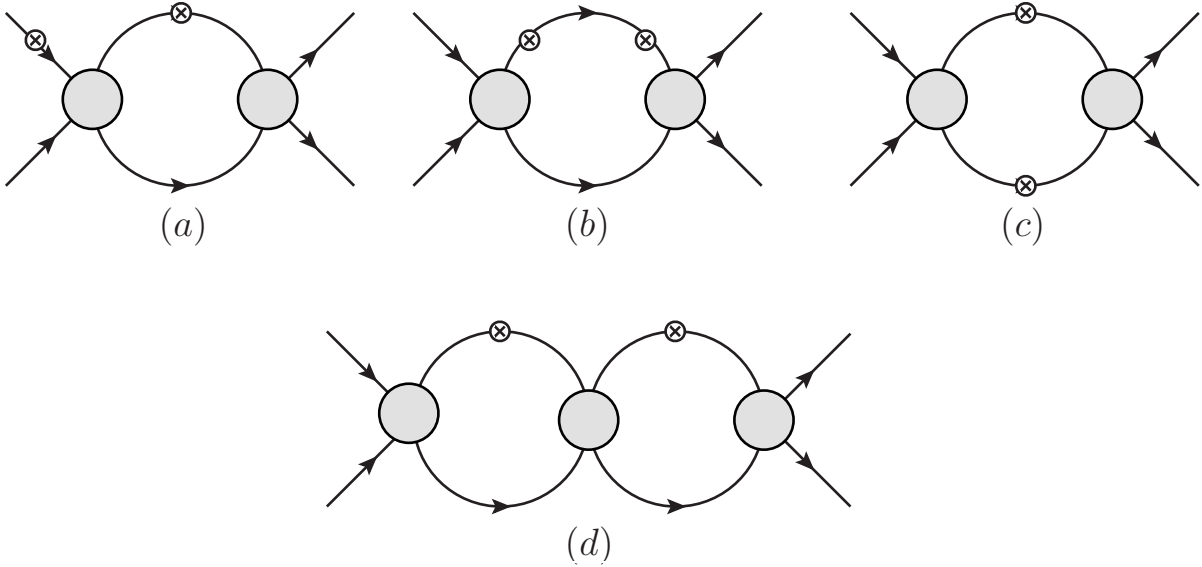
\begin{figure*}[t]
\begin{center}
\fcolorbox{white}{white}{
  \begin{picture}(451,208) (40,-36)
    \SetWidth{1.0}
    \SetColor{Black}
    \Arc[arrow,arrowpos=0.5,arrowlength=5,arrowwidth=2,arrowinset=0.2,clock](106.813,135.125)(33.033,179.256,0.744)
    \Line[arrow,arrowpos=0.5,arrowlength=5,arrowwidth=2,arrowinset=0.2](41.181,168.156)(73.783,135.554)
    \Line[arrow,arrowpos=0.5,arrowlength=5,arrowwidth=2,arrowinset=0.2](41.181,102.953)(73.783,135.554)
    \Arc[arrow,arrowpos=0.5,arrowlength=5,arrowwidth=2,arrowinset=0.2](106.813,135.983)(33.033,-179.256,-0.744)
    \GOval(73.783,135.554)(11.153,11.153)(0){0.882}
    \Line[arrow,arrowpos=0.5,arrowlength=5,arrowwidth=2,arrowinset=0.2](139.844,135.554)(172.446,168.156)
    \Line[arrow,arrowpos=0.5,arrowlength=5,arrowwidth=2,arrowinset=0.2](139.844,135.554)(172.446,102.953)
    \GOval(139.844,135.554)(11.153,11.153)(0){0.882}
    \COval(52.334,157.861)(3.64,3.64)(45.0){Black}{White}\Line(50.514,156.041)(54.154,159.681)\Line(50.514,159.681)(54.154,156.041)
    \COval(107.242,168.156)(3.64,3.64)(45.0){Black}{White}\Line(105.422,166.336)(109.062,169.976)\Line(105.422,169.976)(109.062,166.336)
    \Line[arrow,arrowpos=0.5,arrowlength=5,arrowwidth=2,arrowinset=0.2](293.415,135.554)(326.017,168.156)
    \Arc[arrow,arrowpos=0.5,arrowlength=5,arrowwidth=2,arrowinset=0.2](260.384,135.983)(33.033,-179.256,-0.744)
    \Arc[arrow,arrowpos=0.5,arrowlength=5,arrowwidth=2,arrowinset=0.2,clock](260.384,135.125)(33.033,179.256,0.744)
    \Line[arrow,arrowpos=0.5,arrowlength=5,arrowwidth=2,arrowinset=0.2](194.752,168.156)(227.354,135.554)
    \Line[arrow,arrowpos=0.5,arrowlength=5,arrowwidth=2,arrowinset=0.2](194.752,102.953)(227.354,135.554)
    \Line[arrow,arrowpos=0.5,arrowlength=5,arrowwidth=2,arrowinset=0.2](293.415,135.554)(326.017,102.953)
    \COval(238.507,157.861)(3.64,3.64)(45.0){Black}{White}\Line(236.687,156.041)(240.327,159.681)\Line(236.687,159.681)(240.327,156.041)
    \COval(282.262,157.861)(3.64,3.64)(45.0){Black}{White}\Line(280.442,156.041)(284.082,159.681)\Line(280.442,159.681)(284.082,156.041)
    \GOval(227.354,135.554)(11.153,11.153)(0){0.882}
    \GOval(293.415,135.554)(11.153,11.153)(0){0.882}
    \Arc[arrow,arrowpos=0.5,arrowlength=5,arrowwidth=2,arrowinset=0.2,clock](425.108,135.125)(33.033,179.256,0.744)
    \Arc[arrow,arrowpos=0.5,arrowlength=5,arrowwidth=2,arrowinset=0.2](425.108,135.983)(33.033,-179.256,-0.744)
    \Line[arrow,arrowpos=0.5,arrowlength=5,arrowwidth=2,arrowinset=0.2](359.476,168.156)(392.078,135.554)
    \Line[arrow,arrowpos=0.5,arrowlength=5,arrowwidth=2,arrowinset=0.2](359.476,102.953)(392.078,135.554)
    \Line[arrow,arrowpos=0.5,arrowlength=5,arrowwidth=2,arrowinset=0.2](458.139,135.554)(490.741,168.156)
    \Line[arrow,arrowpos=0.5,arrowlength=5,arrowwidth=2,arrowinset=0.2](458.139,135.554)(490.741,102.953)
    \COval(425.537,168.156)(3.64,3.64)(45.0){Black}{White}\Line(423.717,166.336)(427.357,169.976)\Line(423.717,169.976)(427.357,166.336)
    \COval(425.537,102.953)(3.64,3.64)(45.0){Black}{White}\Line(423.717,101.133)(427.357,104.773)\Line(423.717,104.773)(427.357,101.133)
    \GOval(392.078,135.554)(11.153,11.153)(0){0.882}
    \GOval(458.139,135.554)(11.153,11.153)(0){0.882}
    \Arc[arrow,arrowpos=0.5,arrowlength=5,arrowwidth=2,arrowinset=0.2,clock](227.783,15.014)(33.033,-179.256,-360.744)
    \Arc[arrow,arrowpos=0.5,arrowlength=5,arrowwidth=2,arrowinset=0.2,clock](293.844,15.014)(33.033,-179.256,-360.744)
    \Arc[arrow,arrowpos=0.5,arrowlength=5,arrowwidth=2,arrowinset=0.2](227.783,15.014)(33.033,-179.256,-0.744)
    \Line[arrow,arrowpos=0.5,arrowlength=5,arrowwidth=2,arrowinset=0.2](326.875,14.585)(359.476,47.187)
    \Line[arrow,arrowpos=0.5,arrowlength=5,arrowwidth=2,arrowinset=0.2](326.875,14.585)(359.476,-18.017)
    \Line[arrow,arrowpos=0.5,arrowlength=5,arrowwidth=2,arrowinset=0.2](162.15,48.045)(194.752,15.443)
    \Line[arrow,arrowpos=0.5,arrowlength=5,arrowwidth=2,arrowinset=0.2](162.15,-18.017)(194.752,14.585)
    \Arc[arrow,arrowpos=0.5,arrowlength=5,arrowwidth=2,arrowinset=0.2](293.844,15.014)(33.033,-179.256,-0.744)
    \COval(227.354,48.045)(3.64,3.64)(45.0){Black}{White}\Line(225.534,46.225)(229.174,49.865)\Line(225.534,49.865)(229.174,46.225)
    \COval(293.415,48.045)(3.64,3.64)(45.0){Black}{White}\Line(291.595,46.225)(295.235,49.865)\Line(291.595,49.865)(295.235,46.225)
    \GOval(260.813,14.585)(11.153,11.153)(0){0.882}
    \GOval(193.036,17.159)(11.153,11.153)(0){0.882}
    \GOval(326.875,14.585)(11.153,11.153)(0){0.882}
    \Text(99.242,80.646)[lb]{\Large{\Black{$(a)$}}}
    \Text(252.813,80.646)[lb]{\Large{\Black{$(b)$}}}
    \Text(417.537,80.646)[lb]{\Large{\Black{$(c)$}}}
    \Text(252.813,-40.323)[lb]{\Large{\Black{$(d)$}}}
  \end{picture}
}
\caption{All loop graph topologies contributing to the the $2 \rightarrow 2$ matrix element of the operator $M(q)$.   The shaded blobs denote exact vertices, given in Fig.~\ref{fig:eamp}.  The symbol $\otimes$ denotes an insertion of the operator $\psi^\dagger\psi$.   \label{fig:loop}}
\end{center}
\end{figure*}

The $2\rightarrow2$ matrix element of the LHS of Eq.~(\ref{eq:nnope}) consists of the graphs in Figs.~\ref{fig:tree},\ref{fig:loop}.    In order to match onto operators of dimension $\Delta\leq 5$ in the OPE, it is sufficient to compute the (off-shell) two-particle matrix elements taking the limit of vanishing external energy and momentum, in which case the calculations simplify considerably.

The graphs in Fig.~\ref{fig:tree}, with insertions of $\psi^\dagger\psi$ purely on external lines,  are given by 
\bea
\label{eq:one}
\mbox{Fig.~\ref{fig:tree}}(a) &=& 4 i {\cal A}(q) \left[ {i\over \omega - E_{\vec q}+i\epsilon}\right]^2 + (q\rightarrow -q),\\
\mbox{Fig.~\ref{fig:tree}}(b) &=& 4 i {\cal A}(0)\cdot  {i\over \omega - E_{\vec q}+i\epsilon} \cdot {i\over -\omega - E_{\vec q}+i\epsilon},
\eea
and 
\bea
\label{eq:ossing}
\mbox{Fig.~\ref{fig:tree}}(c)  &=&  4i{\cal A}(0)\cdot  {i\over k^0 - E_{\vec k}+i\epsilon} {i\over  k^0+  \omega - E_{{\vec k}+{\vec q}}+i\epsilon} + (q\rightarrow -q).
\eea
In this last equation $k=(k^0,{\vec k})\rightarrow 0$ is the energy-momentum of any of the external states.    

Note that the graph in Fig.~\ref{fig:tree}(c) are singular when the external states are taken on-shell, $k^0\rightarrow E_{\vec k}$.   To regularize this singularity, we have kept all external states off-shell in the computation of matrix elements of the bilocal operator $M(q)$.      For consistency, we have done this even in the $1\rightarrow 1'$ matching calculation of the previous section.    This is the reason why it is not valid to eliminate the operators in the last line of Eq.~(\ref{eq:1part}) in favor of the quartic operator $\psi_\uparrow^\dagger \psi^\dagger_\downarrow \psi_\uparrow \psi_\downarrow$.    The $1/0$ on-shell singularities in the $2\rightarrow 2'$ matrix element of $M(q)$ are reproduced by similar singularities on the RHS of the OPE (see Eq.~(\ref{eq:singME}) below), and cancel out in the computation of the Wilson coefficients.

Loop corrections to the matrix element are computed by first performing the integration over loop energy by residues.   The result is,
\beq
\mbox{Fig.~\ref{fig:loop}}(a)  = - {4i {\cal A}(0){\cal A}(q)\over \omega - E_{\vec q}+i\epsilon} I_1(q) + (q\rightarrow -q),\\
\eeq
for graphs with a single operator insertion on an internal line.   The graphs with two internal line operator insertions are 
\bea
\mbox{Fig.~\ref{fig:loop}}(b) &=&  {i\over 2} {\cal A}(0)^2  I_2(q) + (q\rightarrow -q),\\
\mbox{Fig.~\ref{fig:loop}}(c) &=&  {i\over 2} {\cal A}(0)^2  I_3(q) + (q\rightarrow -q),\\
\label{eq:last}
\mbox{Fig.~\ref{fig:loop}}(d) &=& - i{\cal A}(0)^2 {\cal A}(q) I^2_1(q) + (q\rightarrow -q).
\eea
In these equations we have introduced
\beq
I_{\alpha=1,2}(q) = \int {d^d {\vec l}\over (2\pi)^d}  \left({1\over E_{\vec l}}\right)^\alpha {1\over \omega - E_{\vec l}-E_{{\vec l}+{\vec q}}+i\epsilon},
\eeq
and,
\beq
I_{3}(q) =\int {d^d {\vec l}\over (2\pi)^d} {1\over E_{\vec l}} {1\over E_{{\vec l}+{\vec q}}} {1\over \omega - E_{\vec l}-E_{{\vec l}+{\vec q}}+i\epsilon}.\\
\eeq
These integrals can be performed by a combination of partial fraction decomposition and Feynman parameter techniques, with the results (with $z=E_{\vec q}/(2\omega +i\epsilon)$)
\begin{widetext}
\bea
I_{1}(q)   &=& -{2  f(\omega,0)\over \omega+i\epsilon} (1-z)^{d/2-2}  {}_2 F_1(1,2-d/2;d/2,z/(1-z)),\\
I_{2}(q)   &=& -{4 f(\omega,0) \over (\omega+i\epsilon)^2} (1-z)^{d/2-3}  {}_2 F_1(2,3-d/2;d/2,z/(1-z)),\\
I_3(q)       &=& {2 I_1(q)\over\omega+i\epsilon}   +  {4m^2\over (4\pi)^{d/2}} {\Gamma^2(d/2-1) \Gamma(2-d/2)\over \Gamma(d-2)} {|{\vec q}|^{d-4}\over \omega+i\epsilon},
\eea
\end{widetext}
where $f(\omega,{\vec q})$ is the function defined in Eq.~(\ref{eq:func}).    

The sum over the above diagrams gives the \emph{exact} matrix element of $\int d^4 x e^{iq\cdot x} T[n(x) n(0)]$ between two particle states.  In order to match this to the OPE, it is necessary to compare this matrix element to the two-particle matrix elements of the RHS of Eq.~(\ref{eq:1part}).   We find that that the matrix element in the $q\rightarrow \infty$ limit is reproduced by the OPE, provided that it includes the quartic operator
\beq
{\cal O}_4 =  \psi_\uparrow^\dagger \psi^\dagger_\downarrow \psi_\uparrow \psi_\downarrow,
\eeq
with a Wilson coefficient given by 
\beq
C_{\psi^4}(q) \langle 2'|{\cal O}_4|2\rangle = \langle 2'|M(q)|2 \rangle  - \langle 2'|\sum_{\alpha'} C_{\alpha'}(q) {\cal O}_{\alpha'}|2\rangle,
\eeq
where in the second line the sum is over the one-body operators appearing in Eqs.~(\ref{eq:cn})-(\ref{eq:c5}).  In dimensional regularization, for off-shell external states, (${\cal O}_{eom}\equiv \psi^\dagger \left(i\partial_t + {\nabla^2\over 2m}\right)\psi$)
\bea
\label{eq:singME}
\langle 2'| n(0)|2 \rangle  &=& 4 i{\cal A}(0) \cdot {i\over k^0-E_{\vec k}},\\
\langle 2'|{\cal O}_{eom}(0)|2 \rangle &=& -4 {\cal A}(0),\\
\label{eq:contME}
\langle 2'|{\cal O}_4(0)|2\rangle &=&-{\cal A}^2(0)/g^2,
\eea
while all other bilinear operators have vanishing two-particle matrix elements at zero external momentum\footnote{More precisely, the matrix elements have power IR divergences for $d>2$ which are set to zero in dimensional regularization.}.    We therefore obtain,
\bea
C_{\psi^4}(q) &=&i g^2\left[{\cal A}(q)\left(I_1(q) - {2\over g} \cdot {1\over \omega - E_{\vec q}}\right)^2 - {1 \over 2} (I_2(q)+I_3(q))+(q\rightarrow-q)\right]\\
& & {}+{4ig}\cdot {1\over \omega - E_{\vec q}}\cdot {1\over \omega +E_{\vec q}}.
\eea
Note that the potential $1/0$ singularities from the on-shell limit of Eq.~(\ref{eq:ossing}) have cancelled against similar divergences in the matrix element of Eq.~(\ref{eq:singME}) leaving behind a finite remainder.   We have also checked, using the method of regions~\cite{Beneke:1997zp}, that all IR divergences in the integrals  $I_{1,2,3}(q)$ cancel, for arbitrary dimension $d$, against similar IR divergences in the two-particle matrix elements of the bilinear operators in Eqs~(\ref{eq:singME})-(\ref{eq:contME}).    Thus the Wilson coefficients are manifestly IR finite as expected on general grounds.

Of the seven operators with $\Delta\leq 5$, only $n(0)$, ${\cal O}_{ij}=\psi^\dagger  \stackrel{\leftrightarrow}{\partial}_i \stackrel{\leftrightarrow}{\partial}_j\psi$, and the two operators on the last line of Eq.~(\ref{eq:1part}), and ${\cal O}_4$ have, by translation and rotational invariance, non-zero values in the many body ground state.  Using rotational/translation invariance, $\langle {\cal O}_{ij}\rangle$ has VEV proportional to the kinetic energy 
\beq
\langle {\cal O}_{ij}\rangle = -{4\over d}\delta_{ij}  \langle |\nabla\psi|^2\rangle,
\eeq
while, by use of the equations of motion
\beq
\langle \psi^\dagger \left(i\partial_t + {\nabla^2\over 2m}\right)\psi\rangle =  \langle\psi^\dagger {\Big(-i\stackrel{\leftarrow}{\partial}_t + {\stackrel{\leftarrow}{\nabla}^2\over 2m}\Big)}\rangle=  -2 g \langle {\cal O}_4\rangle.
\eeq
Thus for $q\rightarrow\infty$, the OPE predicts
\bea
\langle  M(q\rightarrow\infty)  \rangle &\sim & C_n(q) \langle n\rangle -{4\over d} C^{(3)}_{ii}\langle {1\over 2m} |\nabla\psi|^2\rangle + \left[C_{\psi^4}(q) - 2 g C^{(4)}(q) -  2 g C^{(5)}(q)\right] \langle {\cal O}_4\rangle.
\eea
This is more conveniently written in an operator basis consisting of $n(x)$, the Hamiltonian density
\beq
{\cal H} = {1\over 2m} |\nabla\psi|^2 - g(\psi_\uparrow^\dagger \psi_\downarrow^\dagger) (\psi_\uparrow  \psi_\downarrow),
\eeq
and the contact operator ${\cal O}_{c}= g^2  \psi_\uparrow^\dagger \psi^\dagger_\downarrow \psi_\uparrow \psi_\downarrow$.   The result is
 \beq
 \label{eq:vacope}
\langle  M(q)  \rangle = C_n(q) \langle n\rangle + C_{\cal H}(q) \langle {\cal H}\rangle + C_c(q) \langle {\cal O}_c\rangle,
 \eeq
 \vspace{0.1cm}
where $C_n(q)$ is given in Eq.~(\ref{eq:cn}), while 
\beq
\label{eq:ch}
C_{\cal H}(q) = {4i\over d} E_{\vec q}\left[{1\over (\omega - E_{\vec q})^3} -{1\over (\omega +E_{\vec q})^3}\right],
\eeq
and,
\begin{widetext}
\bea
\nn
C_{c}(q) &=& i \left[{\cal A}(q)\left(I_1(q) - {2\over g} \cdot {1\over \omega - E_{\vec q}}\right)^2 -{1 \over 2} I_2(q)-{1\over 2} I_3(q)+(q\rightarrow -q)\right] +{2i\over g}\left[{1\over \omega - E_{\vec q}} + {1\over \omega + E_{\vec q}} \right]^2\\
& & + {4i\over g} {E_{\vec q}\over d} \left[{1\over (\omega - E_{\vec q})^3} -{1\over (\omega +E_{\vec q})^3}\right]
\eea
\end{widetext}
As a consistency check of these results, we note that the Ward identities implied by current conservation, $\partial_t n + \nabla\cdot {\vec J}=0,$ impose the constraint $M(\omega,{\vec q}\rightarrow 0)=0$.   It is straightforward to verify that the Wilson coefficients in Eq.~(\ref{eq:vacope}) all vanish at the point ${\vec q}=0$ as required by the Ward identity.   We also note that the Ward identity requires matrix elements of the operator $n(x)$, and thus $M(q)$ to be renormalization group (RG) invariant.   This property is reflected in our result, involving UV finite (in $d>2$) Wilson coefficients multiplying the RG invariant operators $n$, ${\cal H}$, and ${\cal O}_c$\footnote{RG invariance of ${\cal O}_c$ is derived in ref.~\cite{Braaten:2010if}.}.   

In fact all three operators $n, {\cal H},{\cal O}_c$ appearing in the OPE of $T[n(x) n(0)]$ have a direct thermodynamic interpretation when evaluated in the many-body ground state, in terms of the particle number density $N/V$, the energy density $E/V$ and Tan's contact parameter ${\cal C}$, respectively.   The latter can be expressed as the thermodynamic derivative,
\beq
{\cal C}= \langle {\cal O}_c\rangle = k_B T g^2\partial_g \left.\ln {\cal Z}\right|_{T,\mu}
\eeq
which in $d=3$ can be written in terms of the free energy density ${\cal F}=F/V$
\beq
{\cal C} = {4\pi\over m}\left.{\partial{\cal F}\over \partial a^{-1}}\right|_{T,N}.
\eeq

\subsection{Three spatial dimensions}

We now apply these results to the case $d=3$ of primary physical interest.   In the limit $\omega\rightarrow\infty$, $z=E_{\vec q}/2\omega \ll 1$, the Wilson coefficients can be expanded as
\bea
C_n(q) &=& {i\over\omega} \left({{\vec q}^2\over m \omega}\right) +\cdots,\\
C_{\cal H}(q) &=&  {2i\over\omega^2}\left({ {\vec q}^2\over m\omega}\right)^2+\cdots,
\eea
and, keeping terms up to order $1/g^2$ in the $g\rightarrow\infty$ limit
\bea
\label{eq:exp}
C_c(q) = {4 i \over \omega^2} \left({{\vec q}^2\over m \omega}\right)^2 \left[ {16\over 45}\left(f(\omega)+f(-\omega)\right)+ {2\over 9} \cdot g^{-1} - {i\over 9} \left({1\over f(\omega)}+{1\over f(-\omega)}\right)\cdot g^{-2}\right],
\eea
where $f(\omega) = - m \sqrt{-m(\omega + i\epsilon)}/4\pi$.    Since for real $\omega>0$, $\mbox{Im } G_F(\omega) = \mbox{Im } G_R(\omega),$ we obtain in the long wavelength/large frequency limit
\bea
\label{eq:res1}
S(\omega,{\vec q}) &=&-{m^3\over 4\pi^2} \left({{\vec q}^2\over m \omega}\right)^2 (m\omega)^{-3/2}\left[{16\over 45} + {1\over 9} (a\sqrt{m\omega})^{-2}+\cdots\right]\cdot {\cal C}, 
\eea
which is valid up to corrections to the OPE from operators of dimension $\Delta>5$, as well as corrections from the finite range of the two-body potential (neglected in Eq.~(\ref{eq:lag}).  The $a\rightarrow\infty$ limit of our result agrees with the calculation of~\cite{st}, which uses an equivalent method to ours, but disagrees with that of~\cite{tr}.    The scaling $S(\omega,{\vec q})\sim \omega^{-7/2}$ is consistent with the power counting argument of~\cite{wong} for the case of a contact interaction.

Because $S(\omega,{\vec q})$ falls off as a power for $\omega\rightarrow\infty$, the naive sum rules that follow from expanding the dispersion integrals of Eq.~(\ref{eq:disp}) are not applicable.   Nevertheless, calculating $\mbox{Re} G_F(\omega,{\vec q})$ for real $\omega>0$ and keeping only the leading order term in the OPE, proportional to $\omega^{-2}$, one recovers the usual $f$-sum rule
\beq
\int_0^\infty d\omega\,  \omega S(\omega,{\vec q}) =  {{\vec q}^2\over 2m}\cdot {N\over V}.
\eeq
Perhaps more useful for comparison with experimental data is the Borel sum rule of Eq.~(\ref{eq:borelsr}), which requires evaluation of the Wilson coefficients at imaginary $\omega=i\omega_0$, $(\omega_0>0)$.   From Eq.~(\ref{eq:exp})
\bea
C_c(i\omega_0) = -{im^3\over 4\pi} \left({{\vec q}^2\over m \omega_0}\right)^2 (m\omega_0)^{-3/2}\left[{16\sqrt{2}\over 45}  - {2\over 9} (a\sqrt{m\omega_0})^{-1}-{\sqrt{2}\over 9} (a\sqrt{m\omega_0})^{-2}\right]\cdot {\cal C} +\cdots,
\eea
and therefore, from Eq.~(\ref{eq:borelsr})

\bea
\label{eq:res2}
{1\over \omega_0^2}\int_0^\infty {d\omega^2} e^{-\omega^2/\omega_0^2} S(\omega,{\vec q}) &=& {{\vec q}^2\over m \omega_0^2} \cdot {N\over V}  +  {2\over\omega_0^2} \left({{\vec q}^2\over m \omega_0}\right)^2 \cdot {E\over V} + {m^3\over 4\pi} \left({{\vec q}^2\over m \omega_0}\right)^2 (m\omega_0)^{-3/2}\left[{64\sqrt{2}\over 135\Gamma(3/4)}\right. \\ 
\nn
& & \left.  - {2\over 9} (a\sqrt{m\omega_0})^{-1}-{16\sqrt{2}\over 45\Gamma(1/4)} (a\sqrt{m\omega_0})^{-2}\right]\cdot {\cal C}+\cdots,
\eea
which is the main result of this section.

\section{Polarization Observables}
\label{sec:pol}

The methods of the previous section can also be used to study the asymptotic behavior of the polarized density correlator
\beq
M_{\uparrow\downarrow}(q)=\int d^4 x e^{iq\cdot x} \frac{1}{2}T[n_\uparrow(x) n_\downarrow(0)+n_\downarrow(x) n_\uparrow(0)].
\eeq
In particular, the $1\rightarrow 1$ matrix elements are now identically zero, so that the OPE in this case does not generate bilinear operators.   In the $2\rightarrow 2'$ sector, the matrix elements are given by graphs with the same topologies as in Figs. 3,4.   Matching with zero-momentum, spin singlet external states, the non-zero graphs are
\bea
\mbox{Fig.~\ref{fig:tree}}(a) &=&  i {\cal A}(q) \left[ {i\over \omega - E_{\vec q}+i\epsilon}\right]^2 + (q\rightarrow -q),\\
\mbox{Fig.~\ref{fig:tree}}(b) &=& 2 i {\cal A}(0)\cdot  {i\over \omega - E_{\vec q}+i\epsilon} \cdot {i\over -\omega - E_{\vec q}+i\epsilon},\\
\mbox{Fig.~\ref{fig:loop}}(a)  &=& - {i {\cal A}(0){\cal A}(q)\over \omega - E_{\vec q}+i\epsilon} I_1(q) + (q\rightarrow -q),\\
\mbox{Fig.~\ref{fig:loop}}(c) &=&  {i\over 4} {\cal A}(0)^2  I_3(q) + (q\rightarrow -q),\\
\mbox{Fig.~\ref{fig:loop}}(d) &=& - {i\over 4}{\cal A}(0)^2 {\cal A}(q) I^2_1(q) + (q\rightarrow -q).
\eea
Neglecting operators with $\Delta>5$, the OPE for $M_{\uparrow\downarrow}(q)$ is then
\beq
M_{\uparrow\downarrow}(q)\sim C_c^{\uparrow\downarrow}(q) {\cal O}_c(0),
\eeq
where now
\beq
\label{eq:spinwilson}
C^{\uparrow\downarrow}(q) ={ i\over 4}\left[{\cal A}(q)\left(I_1(q) - {2\over g} \cdot {1\over \omega - E_{\vec q}}\right)^2\ - I_3(q) + (q\rightarrow-q)\right]  + {2i\over g}\cdot {1\over \omega - E_{\vec q}}\cdot {1\over \omega +E_{\vec q}}.
\eeq
Note that due to $SU(2)$ spin symmetry the spin densities $n_\uparrow$, $n_\downarrow$ are separately conserved.    The Ward identities generated by these conservation laws imply, as in the unpolarized case, that $\langle M_{\uparrow\downarrow}(\omega,{\vec q}\rightarrow 0)\rangle \rightarrow 0$.    This is reflected in our result for the OPE, since it follows from Eq.~(\ref{eq:spinwilson}) that $C^{\uparrow\downarrow}(\omega,{\vec q}\rightarrow 0)=0$. 

As before, the asymptotic behavior of the correlator is determined by condensates that encode the thermodynamic properties of the many-body state.   Unlike the unpolarized results of the previous section, the asymptotics is now dominated by the two-body operator ${\cal O}_c$ and therefore the Tan contact parameter ${\cal C}$.

For real $\omega>0$, we find, in the ${\vec q}\rightarrow 0$ limit
\beq
S_{\uparrow\downarrow}(\omega,{\vec q})=-{1\over\pi} \mbox{Im } G^{\uparrow\downarrow}_R(\omega,{\vec q}) \sim -{m^3\over 12 \pi^2} \left({{\vec q}^2\over m\omega}\right) (m\omega)^{-3/2} {\cal C}+ {\cal O}(a^{-2})
\eeq
and thus $S_{\uparrow\downarrow}(\omega,{\vec q}) \sim \omega ^{-5/2}$ in the $\omega\rightarrow\infty$ limit.   In addition, we obtain the sum rule for ${\vec q}\rightarrow 0$ 
\beq
{1\over \omega_0^2}\int_0^\infty {d\omega^2} e^{-\omega^2/\omega_0^2} S^{\uparrow\downarrow}(\omega,{\vec q})  = -{m^3\over 12 \pi} \left({{\vec q}^2\over m\omega_0}\right)\left[{4\sqrt{2}\over\Gamma(1/4)} (m\omega_0)^{-3/2} -{17\over 6} \left({{\vec q}^2\over m\omega_0}\right) (a \sqrt{m\omega_0})^{-1}+\cdots\right]\cdot {\cal C}.
\eeq
Unlike the sum rule for the density structure function in Eq.~(\ref{eq:res2}), this sum rule is dominated by contributions from intermediate two-particle states and is therefore governed by the behavior of the contact $\cal C$ near unitarity.

\section{Bosons near unitarity}
\label{sec:bose}

The results discussed above apply also to bosonic systems near unitarity.   A representative example is He-4, with $a\sim 20 \ell_{vdW}$.  Thus to a good approximation, it is described by the Lagrangian,
\beq
{\cal L} =  {\varphi}^\dagger\left( i\partial_t + {\nabla^2\over 2m}\right)\varphi  - {\lambda\over 4} |\varphi|^4.
\eeq
Then the calculations of the previous sections go through virtually unchanged.   The two-particle scattering amplitude is now
\beq
\label{eq:bamp}
{\cal A}^{-1}(\omega,{\vec q}) = -{1\over \lambda} - {1\over 2} f(\omega,{\vec q})
\eeq
and thus the scattering length in $d=3$ is $a= m\lambda/8\pi$.  In the one-body sector, the OPE of $M(q)$, defined as in Eq.~(\ref{eq:nnope}) with $n(x)=\varphi^\dagger \varphi(x)$ is identical to Eq.~(\ref{eq:1part}) with the replacements $\psi\rightarrow\varphi$.  The $2\rightarrow 2'$ matrix element of $M(q)$ is given by the same graphs as in Figs.~\ref{fig:tree},~\ref{fig:loop}.  These can be obtained from the results of Eqs.~(\ref{eq:one})-(\ref{eq:last}) by inserting a symmetry factor of $1/2$ for every loop, and replacing the amplitude by Eq.~(\ref{eq:bamp}).  Using the equations of motion, 
\beq
\varphi^\dagger (i\partial_t + \nabla^2/2m)\varphi = {\lambda\over 2} |\varphi|^4,
\eeq
the OPE can again be written in the basis of RG invariant operators $n(x)$, the Hamiltonian ${\cal H}(x)$, and the quartic
\beq
{\cal O}_c(x) = {\lambda^2\over 4} |\varphi|^4(x).
\eeq
The result is \footnote{Note that we have dropped the contribution of the dimension-five operator ${\cal H}$.  The reason is that for bosons, there is a three-body operator of the same scaling dimensions which would need to be included as well \cite{braaten2b}.}
 \beq
\langle  M(q\rightarrow\infty)  \rangle \sim  C_n(q) \langle n\rangle 
+ C_c(q) \langle {\cal O}_c\rangle,
 \eeq
 where $C_n(q)$ and $C_{\cal H}(q)$ are given by Eq.~(\ref{eq:cn}) and Eq.~(\ref{eq:ch}) respectively, while $C_c(q)$ now becomes
 \begin{widetext}
\bea
\nn
C_{c}(q) &=& -i \left[{1\over 4}{\cal A}(q)\left(I_1(q) - {4\over \lambda} \cdot {1\over \omega - E_{\vec q}}\right)^2 - {1\over 4} I_2(q)- {1\over 4} I_3(q)+(q\rightarrow -q)\right] -{2i\over \lambda}\left[{1\over \omega - E_{\vec q}} + {1\over \omega + E_{\vec q}} \right]^2\\
& & - {4i\over \lambda} {E_{\vec q}\over d} \left[{1\over (\omega - E_{\vec q})^3} -{1\over (\omega +E_{\vec q})^3}\right].
\eea
\end{widetext}
From these results we obtain the asymptotic behavior of the dynamic structure factor
\bea
\label{eq:hes}
S(\omega,{\vec q}) = {m^3\over 8\pi^2} \left({{\vec q}^2\over m \omega}\right)^2 (m\omega)^{-3/2}\left[{16\over 45}   + {1\over 9} (a\sqrt{m\omega})^{-2}+\cdots\right]\cdot {\cal C}, 
\eea
as well as the sum rule
\bea
\label{eq:he}
{1\over \omega_0^2}\int_0^\infty {d\omega^2} e^{-\omega^2/\omega_0^2} S(\omega,{\vec q}) &=& {{\vec q}^2\over m \omega_0^2} \cdot {N\over V} - {m^3\over 8\pi} \left({{\vec q}^2\over m \omega_0}\right)^2 (m\omega_0)^{-3/2}\left[{64\sqrt{2}\over 135}\Gamma(3/4)\right]
\eea
Note that Eq.~(\ref{eq:hes}) can be compared directly to inelastic neutron scattering data.    In particular, the $\omega^{-7/2}$ tail in $S(\omega,{\vec q})$ is known to provide a good fit to data~\cite{wong}.   We plan to apply the sum rule Eq.~(\ref{eq:he}) against the He-4 data in future work, where in principle one could fit for the value of the contact parameter.

\section{Conclusions and Further Directions}
\label{sec:conc}
In this paper we have used the OPE to develop sum rules for the dynamic structure function. The sum rule relates a weighted integral of $S(\omega,{\vec q})$ to a sum over expectation values of local operators.   There are two sources for corrections to the rule. First of all the sum over operators is truncated in an expansion in powers of $\mu/\omega_0$,  with $\mu$ the chemical potential and $\omega_0$ is a free parameter.   There are also small corrections due to the finite range of the two-body potential, as well as contributions from operators beyond those with $\Delta\leq 5$ considered in this paper.

It would be interesting to use our sum rules to extract the contact parameter ${\cal C}$ in various systems where the finite scattering length universality corrections become non-negligible. Of course, the accuracy of the extraction is limited by the size of finite range effects, which in principle can be included in our sum rules,
at least perturbatively.

In QCD, sum rules similar to the ones presented here have been used to infer properties of low lying
excitations. This is done by assuming that at low energies the spectral density is dominated by a low lying resonance in a particular channel.  The continuum contribution has little effect on the values of the low
energy parameters.  Applying similar ideas to cold atoms systems seems like a natural step.

\acknowledgments
We thank Mark Wise for helpful discussions about off-shell matching and the OPE.   WG is supported by DOE grant DE-FG-02-92ER40704 and by a DOE OJI award.  IR is supported by DOE grant, 22645.1.1110173. IR is thankful to the Caltech high energy theory  group for its hospitality and to the Gordon and Betty Moore foundation for support.

\end{document}